# V1820 Orionis : an RR Lyrae star with strong and irregular Blazhko effect

Pierre de Ponthière [1]
*15 Rue Pré Mathy, Lesve – Profondeville 5170 - Belgium*

Franz-Josef (Josch) Hambsch [1,2,3]
*12 Oude Bleken, Mol, 2400 - Belgium*

Tom Krajci [1]
*P.O. Box 1351, Cloudcroft, NM 88317 - USA*

Kenneth Menzies [1]
*318A Potter Road, Framingham MA,01701 - USA*

Patrick Wils [1,3]
*Aarschotsebaan 31, Hever B-3191, - Belgium*

1 American Association of Variable Star Observers (AAVSO)
2 Bundesdeutsche Arbeitsgemeinschaft für Veränderliche Sterne e.V. (BAV), Germany
3 Vereniging Voor Sterrenkunde (VVS), Belgium

## Abstract

The Blazhko effect in V1820 Orionis and its period were reported for the first time by Wils et al. (2006) from a data analysis of the Northern Sky Variability Survey. The results of additional V1820 Orionis observations over a time span of 4 years are presented herein. From the observed light curves, 73 pulsation maxima have been measured. The times of light maxima have been compared to ephemerides to obtain the (O-C) values. The Blazhko period (27.917 ± 0.002 d) has been derived from light curve Fourier analysis and from ANOVA analyses of the (O-C) values and of magnitudes at maximum light ($M_{max}$). During one Blazhko cycle, a hump in the ascending branch of the light curve was clearly identified and has also created a double maximum in the light curve. The frequency spectrum of the light curve, from a Fourier analysis with Period04, has revealed triplet, quintuplet structures, and a second Blazhko weak modulation (period = 34.72 ±0.01 d). V1820 Orionis can be ranked as a strongly modulated star based on its observed amplitude and phase variations. The amplitude ratio of the largest triplet component to main pulsation component is quite large: 0.34.

## 1. Introduction

The star V1820 Orionis is classified in the General Catalogue of Variable Stars (Samus et al. 2011) as an RR Lyr (RRab) variable star with a period of 0.479067 day and with maximum and minimum magnitudes of 12.5 and 13.4, respectively. This star has been identified as an RRab with a Blazhko period of 28 days by Wils et al. (2006). At this time the star was classified as NSV 02724 in the New Catalogue of Suspected Variable Stars (Kukarkin et al. 2011).

The current data were gathered during 157 nights between December 2006 and March 2012. During this period of 1,918 days, a total of 22,592 magnitude measurements covering 70 Blazhko cycles were collected. The 7,300 observations from December 2006 to February 2010 were made by Hambsch using 30 cm and 50 cm telescopes located in Cloudcroft (New Mexico). From September 2011 to March 2012 the observations were made by Hambsch using the telescope in Cloudcroft and a 40 cm telescope in San Pedro de Atacama (Chile). de Ponthière contributed additional measurements with a 20 cm telescope located at Lesve (Belgium).

The comparison stars used by the authors are given in Table 1. The star coordinates and magnitudes in B and V bands were obtained from the NOMAD catalogue (Zacharias et al. 2011). C1 was used as a magnitude reference and C2 as a check star. The Johnson V magnitudes from different instruments have not been transformed to the standard system since measurements were performed with only a V filter. However, two simultaneous measurements from the instruments in Cloudcroft and San Pedro de Atacama were observed to differ by only 0.025 mag. Dark and flat field corrections were performed with the MAXIMDL software (Diffraction Limited, 2004), aperture photometry was performed using LESVEPHOTOMETRY (de Ponthière, 2010), a custom software which also evaluates the SNR and estimates magnitude errors.

## 2. Light curve maxima analysis

The times of maxima of the light curves have been evaluated with custom software (de Ponthiere, 2010) fitting the light curve with a smoothing spline function (Reinsch, 1967). Table 2 provides the list of observed maxima and Figure 1 shows the (O-C) values.

A linear regression of all available (O-C) values has provided a pulsation period of 0.4790486 d (2.087471 $d^{-1}$). The (O-C) values have been re-evaluated with this new pulsation period. The new pulsation elements derived from a linear least-square fitting are:
$$HJD_{Pulsation} = (2\ 454\ 075.8935 \pm 0.0060) + (0.4790486 \pm 0.0000018)\ E$$
The folded light curve on this pulsation period is shown in Figure 2.

The Blazhko period was determined by a period analysis of the (O–C) values and the $M_{max}$ (Magnitude at Maximum) values with the ANOVA algorithm of PERANSO (Vanmunster, 2007). Both periodograms, presented in Figure 3a and 3b, show a primary peak and a series of aliases equally spaced around the main modulation frequency. The lists of prominent peaks are indicated below.

From (O-C) analysis, the main frequencies and periods are tabulated below:

|       | Frequency (cycles / day) | Period (days) | Peak value |
|-------|--------------------------|---------------|------------|
| $f_0$ | $0.03637 \pm 0.00004$    | $27.50 \pm 0.03$d | 31.8   |
| $f_1$ | 0.03583                  | 27.91         | 27.3       |
| $f_2$ | 0.03471                  | 28.81         | 28.0       |
| $f_3$ | 0.03691                  | 27.09         | 18.2       |
| $f_4$ | 0.03747                  | 26.69         | 20.0       |

From $M_{max}$ analysis, the main frequencies and periods are tabulated below:

|       | Frequency (cycles / day) | Period (days) | Peak value |
|-------|--------------------------|---------------|------------|
| $f_0$ | $0.03581 \pm 0.00004$    | $27.92 \pm 0.03$d | 154.1  |
| $f_1$ | 0.03527                  | 28.35         | 69.6       |
| $f_2$ | 0.03471                  | 28.81         | 61.7       |
| $f_3$ | 0.03637                  | 27.50         | 63.2       |
| $f_4$ | 0.03693                  | 27.08         | 36.3       |

These aliases are apparently due to the presence of two groups of measurements separated by four years. The first group of 13 maxima is centered on January 2007 and the second group of 52 maxima is centered on December 2011. The alias spacing of 0.00054 $d^{-1}$ (1,851 d) is approximately the reciprocal of the time span between the two measurement groups (i.e., 59 months or 1,770 days). A Spectral Window analysis on (O-C) and $M_{max}$ data points provided peaks separated by 0.00057 $d^{-1}$ (1754d), which supports the origin of the aliases.

From the (O-C) period analysis it is not possible to deduce which peak corresponds to the Blazhko period since none of them is emerging significantly. However, the prominent peak of the $M_{max}$ analysis is also found as an important peak in the (O-C) analysis. Therefore, the Blazhko period is estimated as $27.92 \pm 0.03$ days. Wils, et al. (2006) have reported a value of 28 days.

The highest recorded maximum was observed at HJD 2455955.6847, and the Blazhko ephemeris origin has been selected as 69 Blazhko cycles before this highest recorded maximum. On the basis of this origin, the first observations have a positive value for $E_{Blazhko}$.
$$HJD_{Blazhko} = 2454029.2047 + (27.92 \pm 0.03)\ E_{Blazhko}$$

The folded (O-C) and $M_{max}$ versus the Blazhko phase curves are presented in Figures 4 and 5. The magnitudes at maximum values differ by about 0.85 mag, i.e. 52% of the light curve peak to peak variations. If the humps are not taken into account, the (O-C) values differ in a range of 0.075 day, i.e. 15% of the pulsation period.

RR Lyrae stars of type RRab are known to frequently show a hump in their light curves that appears before light maximum (Smith H, 1995). The evolution of a strong hump during five consecutive nights (JD 2455941 to 2455945) was observed. The folded light curves for those nights (JD41 – 45) are given in Figure 6. During the first night (JD 2455941), the shape of the maximum appears normal, but on the second night (JD 2455942) the light curve shows a shoulder in the decreasing branch. On JD 2455943 two maxima separated by 0.0685 day (i.e. 14% of the pulsation period) and on the fourth night, the maximum is preceded by a classical hump in the increasing branch. The (O-C) values for the maxima occurring on nights JD 2455941, JD 2455942 and for the first maximum on night JD 2455943 appear to be outliers in the (O-C) diagram (Figure 1), but the (O-C) value of the shoulder on the second night JD 2455942 is close to what is expected. The evolution of a shockwave phenomenon generating the hump has probably distorted the light curve on JD 2455942 to the point that the magnitude of the hump is larger than at normal maximum magnitude. The same phenomenon is not repeated at each Blazhko cycle, but it probably occurred on nights of JD 2454126 and JD 2454770 as the corresponding (O-C) values appear to be outliers in the (O-C) diagram (Figure 1). These irregularities in the (O-C) values occur around the Blazhko phase equal to 0.5 (Figure 4). The observed shoulder in the decreasing branch on JD 2455942 is similar to the "bump" detected by Jurcsik et al. (2012) in the RZ Lyrae light curve. This phenomenon of light curve distortion for both stars occurs when the Blazhko amplitude modulation is weakest. The bump appearing in the descending branch of RZ Lyrae occurs around pulsation phase 0.25-0.30 but the shoulder in the decreasing branch of V1820 Orionis happens around pulsation phase 0.0. Based on the different pulsation phases at which the light curve distortions occur and the night to night evolution for V1820 Orionis, it can be supposed that the phenomena are probably different for the two stars.

The relationship between $M_{max}$ and (O-C) can be plotted on a diagram. These two quantities vary with the Blazhko phase and if they are repetitive from cycle to cycle, the data will lie on a loop and if they are sinusoidal the loop will be elliptical. The loop will run in a clockwise progression if the $M_{max}$ has positive phase delay versus (O-C) phase and vice-versa. However $M_{max}$ and (O-C) values for V1820 Orionis, represented by small diamonds in Figure 7, are poorly repetitive from cycle to cycle and are largely scattered. The mean values of (O-C) and $M_{max}$ have been evaluated for 10 bins of the Blazhko phase and are represented as large squares in the Figure 7. An inspection of the successive points indicates that the loop is progressing in counter-clockwise direction. The point in the lower left of the diagram with (O-C = -0.1 day) corresponds to the strong hump described above. Le Borgne JF, et al. (2012) have shown that for most of the analyzed Blazhko stars the $M_{max}$ versus (O-C) diagrams exhibit a similar counter-clockwise rotation

## 3. Frequency spectrum analysis

A Blazhko effect on the light-curve can be modeled as an amplitude and/or phase modulation of the periodic pulsation, with the reciprocal of the modulation frequency being the Blazhko period. Szeidl B. & Jurcsik J. (2009) have shown that the Fourier spectrum of an amplitude and phase modulation model is given by an infinite series including the fundamental frequency ($f_0$), harmonic frequencies ($if_0$) and multiplet frequencies ($if_0 \pm jf_B$)

$$m(t) = \sum A_i \sin(i\omega t + \Phi_{i0}) + \sum\sum A^+_{ij} \sin[(i\omega + j\Omega)t + \Phi^+_{ij}] + \sum\sum A^-_{ij} \sin[(i\omega - j\Omega)t + \Phi^-_{ij}]$$

where:
    $\omega = 2\pi f_0$, $f_0$ is the fundamental frequency of the light-curve,
    $\Omega = 2\pi f_B$, $f_B$ is the Blazhko modulation frequency, and
    $A_i$, $A^+_{ij}$, $A^-_{ij}$ are the Fourier coefficients and $\Phi_i$, $\Phi^+_{ij}$ and $\Phi^-_{ijj}$ their phase angles, "i" indices are used for the fundamental and the harmonic frequencies and "j" indices for the side lobes (e.g. $A^+_{32}$ is the coefficient of the multiplet ($3f_0 + 2f_B$)).

The methodology used herein is similar to the one reported by Kolenberg K. (2009) where triplet and quintuplet components were detected in the spectrum of SS Fornacis. The spectral analysis was performed with Period04 (Lenz & Breger, 2005) to yield a Fourier analysis and multi-frequency sine-wave fitting.

The sine-wave fitting was determined by successive data pre-whitening and Fourier analysis on residuals. For each observed harmonic and triplet, the signal to noise ratio has been evaluated to retain only significant signals, i.e. with an SNR greater than 3.5. During the Period04 sine-wave fitting process, only the fundamental $f_0$ and the first main triplet component $f_0 + f_B$ frequencies have been unconstrained, the other frequencies have been entered as combinations of $f_0$ and $f_0 + f_B$. Table 3 provides the amplitude and phase for each Fourier component obtained with the best sine-wave fitting. The uncertainties of frequencies, amplitudes and phases have been estimated by Monte Carlo simulations. As it is known that Monte Carlo simulation uncertainties can be underestimated (Kolenberg et al. 2009), the uncertainty values have been multiplied by a factor of two. The harmonics of $f_0$ are significant to the $8^{th}$ order. The residuals after subtraction of the best fit based on $f_0$ and harmonics up to the $8^{th}$ order is provided in Figure 8a. The large residuals close to the phase of maximum light (0.8 to 1.1) are due to amplitude and phase modulations created by the Blazhko effect. The residuals are reduced significantly when the side peaks (triplets) around the $f_0$ frequency and harmonics up to the $7^{th}$ order are included in the fitting process (Figure 8b).

The fundamental pulsation frequency $f_0$ (2.08747 d$^{-1}$) is very close to the frequency obtained from a linear regression analysis of time of maxima. The Blazhko period was also measured from the first side peak frequency $f_0 + f_B$ and $f_0$ to yield $f_B$ = (2.12329– 2.08747) = 0.03582 d$^{-1}$ and $P_B$ = 27.917 ± 0.002 days. The second side peak frequency $f_0 - f_B$ exhibited a lower amplitude and higher uncertainty and was not used for Blazhko period evaluation. The Blazhko period found with the sine-wave fitting method (27.919 ± 0.002 days) is equal to the value found with the brightness at maximum analysis (27.92 ± 0.03 days). Table 4 lists the harmonic and significant amplitude ratios. One useful parameter to quantify the Blazhko effect is the amplitude ratio $A^{\pm}_{11} / A_1$, where $A^{\pm}_{11}$ is the amplitude of the largest side lobe at $f_0+f_B$ or $f_0-f_B$ and $A_1$ is the amplitude of $f_0$. The most common and maximum values for this ratio are 0.15 and 0.4, respectively (Alcock C. et al 2003). With an amplitude ratio $A_{11}^+/A_1$= 0.34, V1820 Orionis can be ranked as strongly modulated.

The triplet ratios $R_i = A^+_{i1} / A^-_{i1}$ and asymmetries $Q_i = (A^+_{i1} - A^-_{i1}) / (A^+_{i1} + A^-_{i1})$ are also provided in Table 4. The asymmetry in the side lobes observed for V1820 Orionis is not unexpected on the basis of Szeidl B. & Jurcsik J. (2009), which showed that this asymmetry is related to the phase difference between the Blazhko amplitude and phase modulations. If the Blazhko effect was limited to amplitude modulation, the ratios $R_i$ and $Q_i$ would be equal to 1 and 0, respectively. The asymmetry ratios $Q_i$ around 0.35 are a sign that V1820 Orionis is amplitude and phase modulated.

Besides the harmonics and triplets, some quintuplet components ($kf_0 + 2f_B$) and a peak at the Blazhko period itself, were found. And finally, two modulation peaks appear in the spectrum around $f_0$ and $2f_0$ with a separation of 0.028 d$^{-1}$. They correspond to a second Blazhko modulation $f_{B2}$ (1/$f_{B2}$ = 34.72 ±0.01 days). This phenomenon of multi-periodic modulation has also been detected by Sódor et al. (2011) in the spectrum of CZ Lacertae. In the case of CZ Lacertae, the modulation components of the two frequencies ($f_B$ and $f_{B2}$) have similar amplitudes, which is not the case for V1820 Orionis. The second modulation frequency $f_{B2}$ has weaker components than $f_B$, but they remain significant as their SNR are 9.2 and 6.0, respectively. A spectral analysis on $M_{max}$ values provides the same two Blazhko modulation frequencies $f_B$ and $f_{B2}$ which are in a 5:4 resonance ratio. The corresponding beating period is 139 days which is visible on the multi-frequency sine-wave fitting obtained with Period04 (Figure 9). For clarity, Figure 9 only includes the last observation season (2011-2012). A 5:4 resonance ratio was also found between the two modulation frequencies of CZ Lacertae during the 2004 observation season (Sódor et al. 2011) but the next year this resonance ratio changed to a value of 4:3.

## 4. Light curve variations over Blazhko cycle

In order to investigate the light curve variations over the Blazhko cycle, the complete dataset was subdivided into 10 temporal subsets corresponding to different Blazhko phase intervals $\Delta\Psi_i$ (i=0,9). The ephemeris derived previously during the analysis of light curve maxima was used to define the epoch of the Blazhko zero phase (HJD = 2454029.2047). The data points are relatively well distributed over the subsets with the number of data points in each subset varying between 1,144 and 3,045. The light curves for each subset are presented in Figure 10. The strong hump observed during JD 2455942, as described previously, is highlighted in red in the panel of subset ($\Delta\Psi$ = 0.5 – 0.6).

For each subset, the amplitude $A_i$ and phase $\Phi_i$ of of the fundamental and harmonic frequencies up to the 4$^{th}$ order have been evaluated with the Least-Square Fit module of Period04. The amplitudes and epoch-independent phase differences ($\Phi_{k1} = \Phi_k - k\Phi_1$) over the Blazhko cycle are provided in Table 4 and exhibited in Figures 11a and 11b. As expected, the amplitude of the fundamental frequency is clearly lower at a Blazhko phase around 0.5, i.e. when the light curve amplitude variation on the pulsation cycle is weaker. The maximum and minimum $\Phi_1$ phase values (2.234 and 1.253 radians) are found in subsets $\Delta\Psi$ (0.4-0.5) and $\Delta\Psi$ (0.1-0.2), respectively. The difference between maximum and minimum $\Phi_1$ phase is a measure of the phase modulation strength and is equal to (2.234 - 1.253) = 0.981 radian or 0.156 cycle, which corresponds roughly to the value of 15 % noted for the peak to peak deviation of (O-C). The phase variation of harmonic component $\Phi_{41}$ is the largest with a value of 3.6 radians while $\Phi_{21}$ varies only by 0.37 radians over the Blazhko cycle. The phase variation of harmonic component $\Phi_{41}$ is very large as compared to $\Phi_{k1}$ values of 1 and 0.5 radians observed for RZ Lyrae (Jurcsik et al. 2012) and SS Fornacis (Kolenberg et al. 2009), respectively.

Figure 12 provides a graph of the $A_1$ coefficient versus the $\Phi_1$ for the different subsets. This graph is similar to the graph of the Magnitude at Maximum versus (O-C) given in Figure 7. In order to compare the two graphs, the $\Phi_1$ axis of Figure 12 was inverted. Indeed for the sine-wave fitting [$A_i \sin(\omega_i t + \Phi_i)$], a larger $\Phi_i$ phase corresponds to a time advance, i.e. a lower value of (O-C). With the $\Phi_1$ axis inverted, the loop of $A_1$ values exhibits a counter-clockwise progression as in Figure 7.

## 5. Conclusions

The strong and irregular Blazhko behavior of V1820 Orionis has been exhibited by two different methods: (1) measurement of light curve maxima and (2) Fourier analysis. The later method, Fourier analysis, has been feasible due to the large number of regular observations over the pulsation period which was not limited to the times of light curve maxima. Both methods yield the same results for the fundamental pulsation period (0.4790486 day ± 0.0000018) and the Blazhko period, (27.917 days ± 0.002). The irregularities of the Blazhko effect are probably explained by variations of the strength of a shockwave phenomenon generating the hump in the ascending branch of the light curve. This erratic behavior generally occurs around a Blazhko phase of 0.5. Measured ratios of Fourier amplitudes and their asymmetries also confirm strong Blazhko amplitude and phase modulations. A second weaker Blazhko modulation with a period 34.70 ±0.02 days has also been identified. The two modulation frequencies are in a 5:4 resonance ratio.

## Acknowledgements

Dr A. Henden, Director of AAVSO and the AAVSO are acknowledged for the use of AAVSOnet telescopes at Cloudcroft (New Mexico, USA). The authors thank Dr K. Kolenberg for helpful suggestions in the Fourier analysis and the referee Dr J. Jurcsik for constructive comments which have helped to clarify and improve the paper.

### Table 1. Comparison stars for V1820 Orionis

| Identification | R.A. (2000) h m s | Dec (2000) ° ' " | B | V | B-V | |
|---|---|---|---|---|---|---|
| GSC 125-41 | 05 54 57.4 | +04 56 42.5 | 13.83 | 13.31 | 0.52 | C1 |
| GSC 125-341 | 05 54 29.6 | +04 53 59.1 | 14.94 | 13.65 | 1.29 | C2 |

### Table 2. List of measured maxima of V1820 Orionis

| Maximum HJD | Error | O-C (day) | E | Magnitude | Error | Filter | Location | Remark |
|---|---|---|---|---|---|---|---|---|
| 2454075.9173 | 0.0033 | 0.0235 | 0 | 12.367 | 0.007 | C | Cloudcroft | |
| 2454076.8783 | 0.0035 | 0.0264 | 2 | 12.330 | 0.006 | C | Cloudcroft | |
| 2454079.7429 | 0.0020 | 0.0167 | 8 | 12.083 | 0.005 | C | Cloudcroft | |
| 2454085.9515 | 0.0011 | -0.0024 | 21 | 11.851 | 0.005 | C | Cloudcroft | |
| 2454104.6673 | 0.0028 | 0.0306 | 60 | 12.342 | 0.005 | C | Cloudcroft | |
| 2454110.8696 | 0.0007 | 0.0053 | 73 | 11.902 | 0.004 | C | Cloudcroft | |
| 2454114.6902 | 0.0008 | -0.0065 | 81 | 11.844 | 0.003 | C | Cloudcroft | |
| 2454126.6093 | 0.0023 | -0.0636 | 106 | 12.393 | 0.004 | C | Cloudcroft | hump |
| 2454135.7878 | 0.0015 | 0.0130 | 125 | 12.158 | 0.004 | C | Cloudcroft | |
| 2454136.7440 | 0.0018 | 0.0111 | 127 | 12.122 | 0.004 | C | Cloudcroft | |
| 2454137.6982 | 0.0009 | 0.0072 | 129 | 12.021 | 0.003 | C | Cloudcroft | |
| 2454149.6405 | 0.0012 | -0.0267 | 154 | 12.118 | 0.004 | C | Cloudcroft | |
| 2454162.6332 | 0.0023 | 0.0317 | 181 | 12.299 | 0.004 | C | Cloudcroft | |
| 2454439.9682 | 0.0039 | -0.00205 | 760 | 12.348 | 0.005 | V | Cloudcroft | |
| 2454748.9571 | 0.0035 | 0.00096 | 1405 | 12.132 | 0.007 | V | Cloudcroft | |
| 2454749.9218 | 0.0053 | 0.00757 | 1407 | 12.046 | 0.043 | V | Cloudcroft | |
| 2454770.8990 | 0.0036 | -0.09334 | 1451 | 12.370 | 0.021 | C | Cloudcroft | hump |
| 2454832.7977 | 0.0025 | 0.00818 | 1580 | 12.288 | 0.009 | V | Cloudcroft | |
| 2455245.7039 | 0.0019 | -0.02490 | 2442 | 12.356 | 0.011 | V | Cloudcroft | |
| 2455813.8853 | 0.0011 | 0.00570 | 3628 | 11.827 | 0.008 | V | Chile | |
| 2455824.8744 | 0.0027 | -0.02330 | 3651 | 12.203 | 0.010 | V | Chile | |
| 2455825.8249 | 0.0034 | -0.03090 | 3653 | 12.242 | 0.009 | V | Chile | |
| 2455845.9745 | 0.0025 | -0.00131 | 3695 | 11.798 | 0.016 | V | Chile | |
| 2455859.8570 | 0.0033 | -0.01120 | 3724 | 12.428 | 0.009 | V | Chile | |
| 2455861.7776 | 0.0045 | -0.00679 | 3728 | 12.444 | 0.009 | V | Chile | |
| 2455871.8475 | 0.0023 | 0.00310 | 3749 | 11.870 | 0.017 | V | Chile | |
| 2455872.8024 | 0.0017 | -0.00009 | 3751 | 11.886 | 0.008 | V | Chile | |
| 2455873.7629 | 0.0018 | 0.00231 | 3753 | 11.923 | 0.010 | V | Chile | |
| 2455881.8771 | 0.0048 | -0.02730 | 3770 | 12.225 | 0.018 | V | Chile | |
| 2455883.8157 | 0.0026 | -0.00490 | 3774 | 12.281 | 0.010 | V | Chile | |
| 2455884.7742 | 0.0037 | -0.00449 | 3776 | 12.288 | 0.010 | V | Chile | |
| 2455885.7271 | 0.0035 | -0.00969 | 3778 | 12.298 | 0.009 | V | Chile | |
| 2455886.6856 | 0.0027 | -0.00928 | 3780 | 12.325 | 0.009 | V | Chile | |
| 2455887.6455 | 0.0032 | -0.00748 | 3782 | 12.326 | 0.009 | V | Chile | |
| 2455894.8573 | 0.0022 | 0.01860 | 3797 | 12.137 | 0.010 | V | Chile | |
| 2455895.8119 | 0.0023 | 0.01511 | 3799 | 12.080 | 0.010 | V | Chile | |
| 2455896.7562 | 0.0016 | 0.00131 | 3801 | 11.981 | 0.020 | V | Chile | |
| 2455896.7649 | 0.0020 | 0.01001 | 3801 | 11.924 | 0.009 | V | Chile | |

| | | | | | | | |
|---|---|---|---|---|---|---|---|
| 2455897.7287 | 0.0028 | 0.01572 | 3803 | 11.998 | 0.009 | V | Chile |
| 2455898.6824 | 0.0016 | 0.01132 | 3805 | 11.955 | 0.009 | V | Chile |
| 2455899.6376 | 0.0015 | 0.00842 | 3807 | 11.970 | 0.010 | V | Chile |
| 2455900.5927 | 0.0028 | 0.00543 | 3809 | 11.977 | 0.014 | V | Chile |
| 2455905.8498 | 0.0051 | -0.00700 | 3820 | 12.129 | 0.026 | V | Chile |
| 2455906.8131 | 0.0045 | -0.00179 | 3822 | 12.171 | 0.032 | V | Chile |
| 2455907.7555 | 0.0041 | -0.01749 | 3824 | 12.189 | 0.027 | V | Chile |
| 2455908.7226 | 0.0060 | -0.00849 | 3826 | 12.269 | 0.032 | V | Chile |
| 2455909.6812 | 0.0063 | -0.00798 | 3828 | 12.373 | 0.014 | V | Chile |
| 2455910.6170 | 0.0051 | -0.03028 | 3830 | 12.339 | 0.012 | V | Chile |
| 2455919.7662 | 0.0033 | 0.01701 | 3849 | 12.220 | 0.014 | V | Chile |
| 2455930.7657 | 0.0041 | -0.00159 | 3872 | 11.928 | 0.023 | V | Cloudcroft |
| 2455931.7210 | 0.0024 | -0.00438 | 3874 | 11.965 | 0.025 | V | Cloudcroft |
| 2455932.6809 | 0.0024 | -0.00258 | 3876 | 12.074 | 0.019 | V | Cloudcroft |
| 2455941.7342 | 0.0050 | -0.05119 | 3895 | 12.461 | 0.010 | V | Chile  hump |
| 2455942.6755 | 0.0028 | -0.06799 | 3897 | 12.399 | 0.009 | V | Chile  hump |
| 2455943.6482 | 0.0064 | -0.05338 | 3899 | 12.502 | 0.010 | V | Chile  hump |
| 2455943.7167 | 0.0074 | 0.01512 | 3899 | 12.533 | 0.010 | V | Chile |
| 2455944.6666 | 0.0084 | 0.00692 | 3901 | 12.514 | 0.009 | V | Chile |
| 2455945.6294 | 0.0025 | 0.01163 | 3903 | 12.466 | 0.016 | V | Cloudcroft |
| 2455945.6433 | 0.0072 | 0.02553 | 3903 | 12.446 | 0.009 | V | Chile |
| 2455946.6153 | 0.0072 | 0.03943 | 3905 | 12.442 | 0.018 | V | Cloudcroft |
| 2455946.6215 | 0.0066 | 0.04563 | 3905 | 12.411 | 0.009 | V | Chile |
| 2455947.5790 | 0.0064 | 0.04504 | 3907 | 12.347 | 0.010 | V | Chile |
| 2455948.5257 | 0.0031 | 0.03364 | 3909 | 12.228 | 0.010 | V | Chile |
| 2455953.7740 | 0.0017 | 0.01241 | 3920 | 11.822 | 0.015 | V | Cloudcroft |
| 2455954.2495 | 0.0026 | 0.00886 | 3921 | 11.838 | 0.015 | V | Lesve |
| 2455955.6847 | 0.0015 | 0.00692 | 3924 | 11.681 | 0.014 | V | Chile |
| 2455956.6423 | 0.0012 | 0.00643 | 3926 | 11.734 | 0.013 | V | Cloudcroft |
| 2455957.5987 | 0.0013 | 0.00473 | 3928 | 11.729 | 0.020 | V | Cloudcroft |
| 2455964.2835 | 0.0037 | -0.01714 | 3942 | 12.284 | 0.014 | V | Lesve |
| 2455978.6990 | 0.0030 | 0.02692 | 3972 | 12.177 | 0.009 | V | Cloudcroft |
| 2455979.6582 | 0.0022 | 0.02803 | 3974 | 12.066 | 0.009 | V | Cloudcroft |
| 2455980.6063 | 0.0012 | 0.01803 | 3976 | 11.953 | 0.009 | V | Cloudcroft |
| 2455989.6865 | 0.0029 | -0.00368 | 3995 | 12.018 | 0.020 | V | Cloudcroft |

**Table 3. Multi-frequency fit results**
*The frequency uncertainties on $f_0$, $f_0 + f_B$ and $f_0 + f_{B2}$ are $4\ 10^{-7}$, $2\ 10^{-6}$, $7\ 10^{-6}$ respectively.*
*The values displayed in italics correspond to components not exceeding a SNR greater than 3.5*

| | $f\ [d^{-1}]$ | $A_i$ [mag] ±0.002 | $\Phi$[cycle] | $\sigma(\Phi)$ | SNR |
|---|---|---|---|---|---|
| $f_o$ | 2.087466 | 0.335 | 0.245768 | 0.001 | 82.1 |
| $2f_o$ | 4.174932 | 0.126 | 0.851262 | 0.002 | 34.8 |
| $3f_o$ | 6.262398 | 0.062 | 0.459455 | 0.003 | 19.8 |
| $4f_o$ | 8.349864 | 0.031 | 0.070836 | 0.006 | 11.6 |
| $5f_o$ | 10.43733 | 0.020 | 0.703054 | 0.010 | 8.3 |
| $6f_o$ | 12.5248 | 0.014 | 0.296387 | 0.014 | 6.1 |
| $7f_o$ | 14.61226 | 0.011 | 0.914742 | 0.016 | 5.4 |
| $8f_o$ | 16.69973 | *0.007* | *0.506391* | *0.028* | *3.3* |
| $f_B$ | 0.035824 | 0.026 | 0.113932 | 0.007 | 5.8 |
| $f_o + f_B$ | 2.12329 | 0.114 | 0.08408 | 0.002 | 28.0 |
| $f_o - f_B$ | 2.051642 | 0.056 | 0.16969 | 0.003 | 13.8 |
| $2f_o + f_B$ | 4.210756 | 0.078 | 0.672295 | 0.003 | 21.8 |
| $2f_o - f_B$ | 4.139108 | 0.040 | 0.756026 | 0.005 | 11.2 |
| $3f_o + f_B$ | 6.298222 | 0.065 | 0.335476 | 0.003 | 20.7 |
| $3f_o - f_B$ | 6.226574 | 0.032 | 0.435456 | 0.006 | 10.2 |
| $4f_o + f_B$ | 8.385687 | 0.041 | 0.972183 | 0.005 | 15.4 |
| $4f_o - f_B$ | 8.31404 | 0.019 | 0.089832 | 0.010 | 7.2 |
| $5f_o + f_B$ | 10.47315 | 0.024 | 0.599442 | 0.008 | 9.9 |
| $5f_o - f_B$ | 10.40151 | 0.011 | 0.722088 | 0.017 | 4.8 |
| $6f_o + f_B$ | 12.56062 | 0.016 | 0.183182 | 0.012 | 7.1 |
| $6f_o - f_B$ | 12.48897 | *0.007* | *0.299414* | *0.027* | *3.2* |
| $7f_o + f_B$ | 14.64809 | 0.012 | 0.783367 | 0.016 | 5.6 |
| $f_o + 2f_B$ | 2.159114 | *0.014* | *0.494921* | *0.015* | *3.3* |
| $3f_o + 2f_B$ | 6.334046 | 0.021 | 0.081559 | 0.009 | 6.7 |
| $4f_o + 2f_B$ | 8.421511 | 0.023 | 0.818861 | 0.009 | 8.5 |
| $f_o + f_{B2}$ | 2.116266 | 0.028 | 0.90776 | 0.007 | 7.0 |
| $2f_o + f_{B2}$ | 4.203732 | 0.024 | 0.509457 | 0.008 | 6.7 |

**Table 4. Harmonic, Triplet amplitudes, ratios and asymmetry parameters**

| k | $A_i/A_1$ | $A^+_{i1}/A_1$ | $A^-_{i1}/A_1$ | $R_i$ | $Q_i$ |
|---|---|---|---|---|---|
| 1 | 1.00 | 0.34 | 0.17 | 2.01 | 0.34 |
| 2 | 0.37 | 0.23 | 0.12 | 1.96 | 0.32 |
| 3 | 0.18 | 0.19 | 0.10 | 2.02 | 0.34 |
| 4 | 0.09 | 0.12 | 0.06 | 2.14 | 0.36 |
| 5 | 0.06 | 0.07 | 0.03 | 2.08 | 0.35 |
| 6 | 0.04 | 0.05 | 0.02 | 2.20 | 0.37 |
| 7 | 0.03 | | | | |
| 8 | 0.02 | | | | |

**Table 5. Fourier coefficients over Blazhko cycle**

| Ψ (cycle) | $A_1$ (mag) | $A_2$ (mag) | $A_3$ (mag) | $A_4$ (mag) | $\Phi_1$ (rad) | $\Phi_{21}$ (rad) | $\Phi_{31}$ (rad) | $\Phi_{41}$ (rad) |
|---|---|---|---|---|---|---|---|---|
| 0.0 - 0.1 | 0.466 | 0.223 | 0.149 | 0.108 | 1.411 | 2.225 | 4.642 | 0.873 |
| 0.1 - 0.2 | 0.432 | 0.188 | 0.134 | 0.070 | 1.253 | 2.134 | 4.667 | 0.910 |
| 0.2 - 0.3 | 0.367 | 0.156 | 0.100 | 0.039 | 1.368 | 2.219 | 4.604 | 0.814 |
| 0.3 - 0.4 | 0.320 | 0.150 | 0.080 | 0.032 | 1.752 | 2.269 | 4.897 | 1.293 |
| 0.4 - 0.5 | 0.240 | 0.100 | 0.028 | 0.016 | 2.234 | 2.180 | 5.353 | 2.206 |
| 0.5 - 0.6 | 0.219 | 0.094 | 0.042 | 0.016 | 2.227 | 2.323 | 5.948 | 4.482 |
| 0.6 - 0.7 | 0.231 | 0.055 | 0.018 | 0.016 | 1.567 | 2.485 | 5.091 | 2.699 |
| 0.7 - 0.8 | 0.326 | 0.112 | 0.047 | 0.014 | 1.344 | 2.503 | 5.267 | 2.544 |
| 0.8 - 0.9 | 0.401 | 0.172 | 0.098 | 0.055 | 1.266 | 2.269 | 4.809 | 1.052 |
| 0.9 - 1.0 | 0.462 | 0.211 | 0.149 | 0.099 | 1.275 | 2.362 | 4.777 | 1.051 |

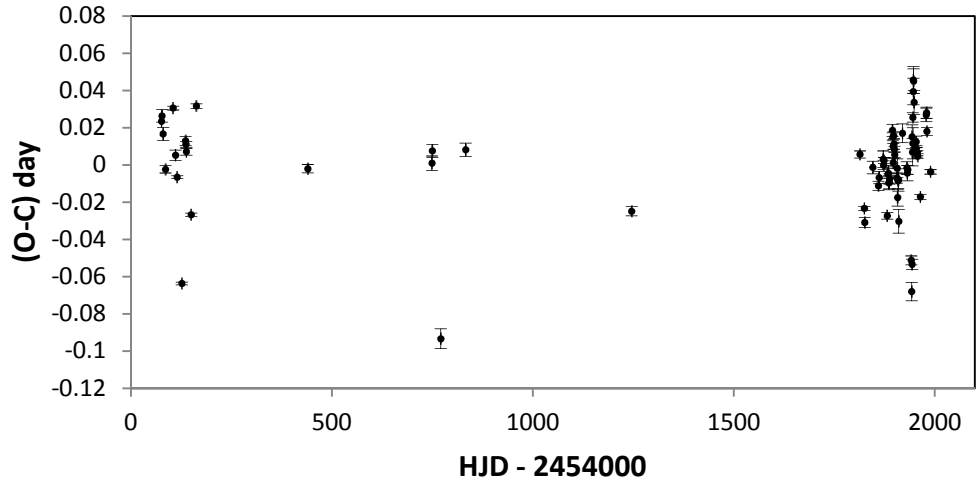

Figure 1. V1820 Orionis (O-C)

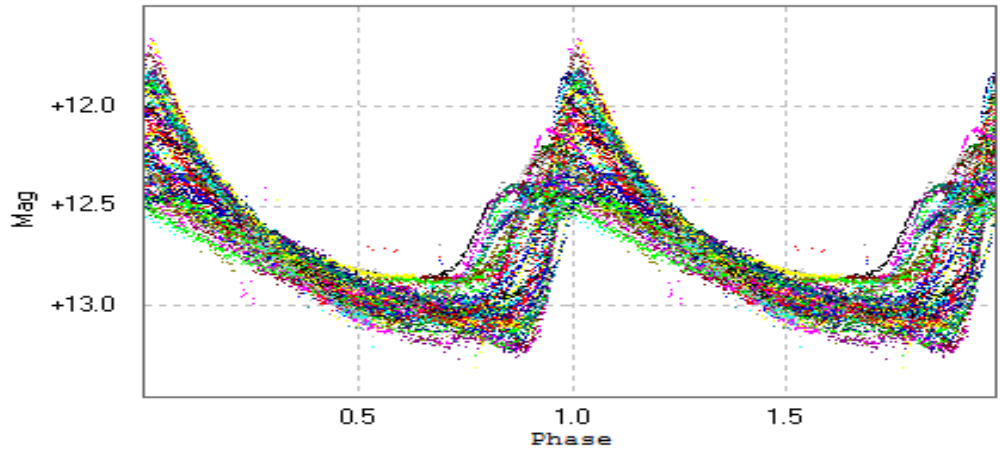
Figure 2. Light curve folded with pulsation period

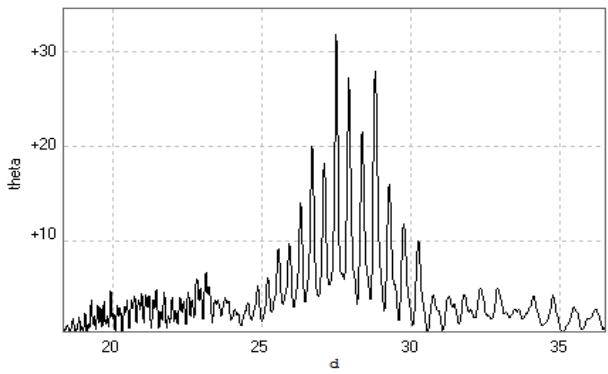
Figure 3a. (O-C) Periodogram

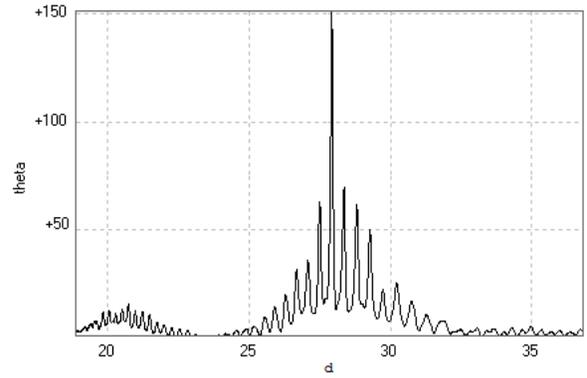
Figure 3b. Magnitude at maximum Periodogram

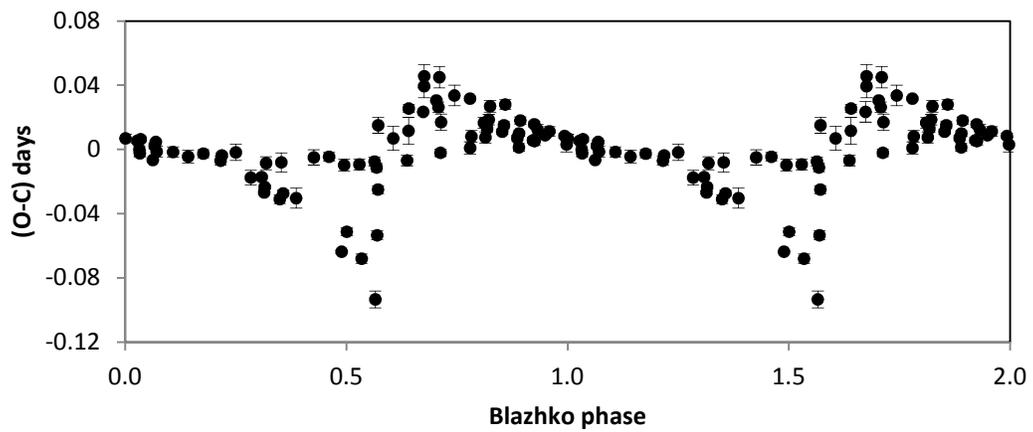
Figure 4. O-C [days] versus Blazhko phase

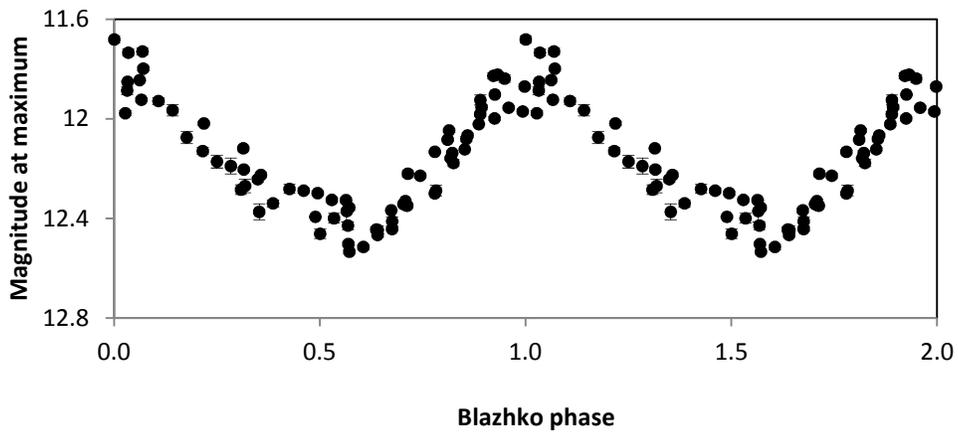

Figure 5. Magnitude [mag] at maximum versus Blazhko phase

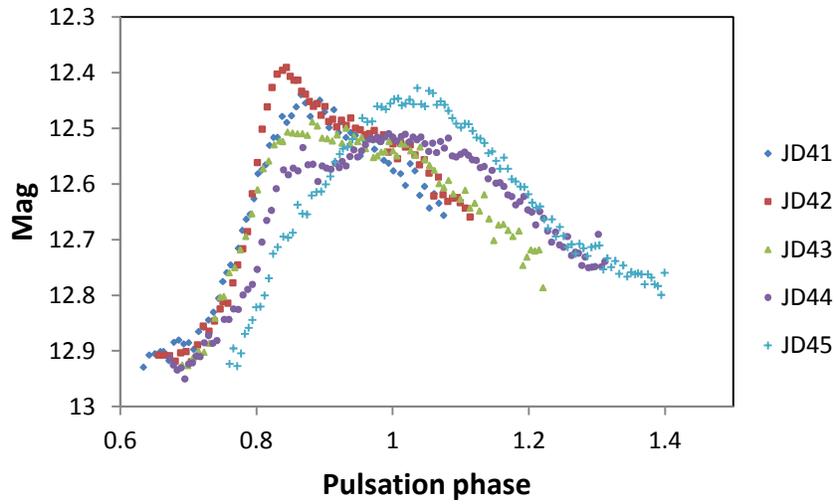

Fig. 6 Hump evolution for nights JD 2455941 to 2455945

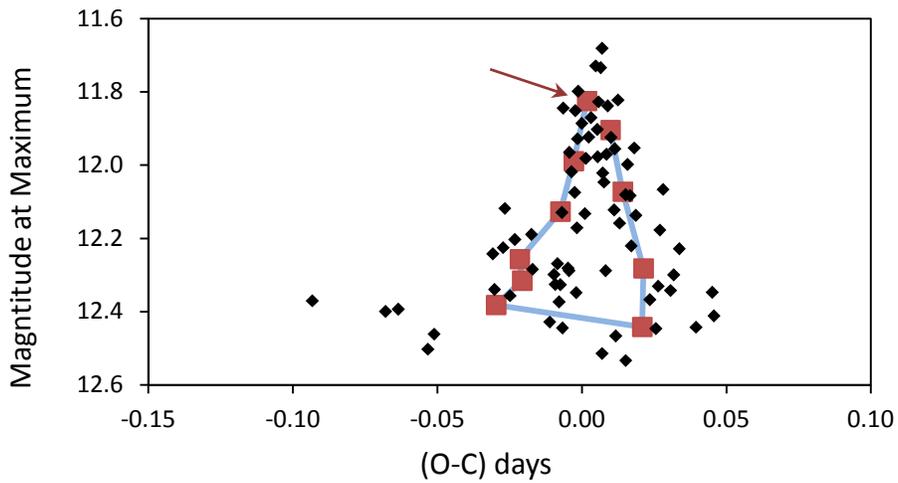

Figure 7. Magnitude at Maximum versus O-C [days] values.
*Individual values and their means are represented as small diamonds and large squares respectively.*
*The point corresponding to the bin nearest to 0.0 Blazhko phase is indicated with a small arrow.*

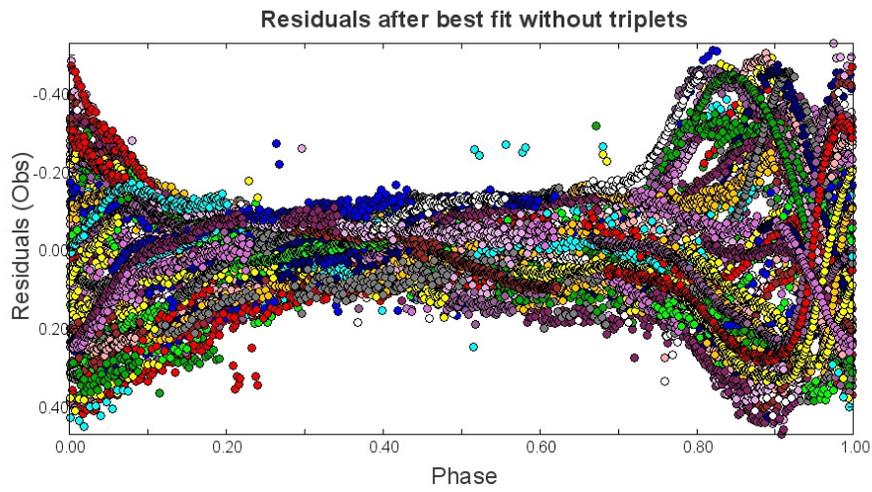

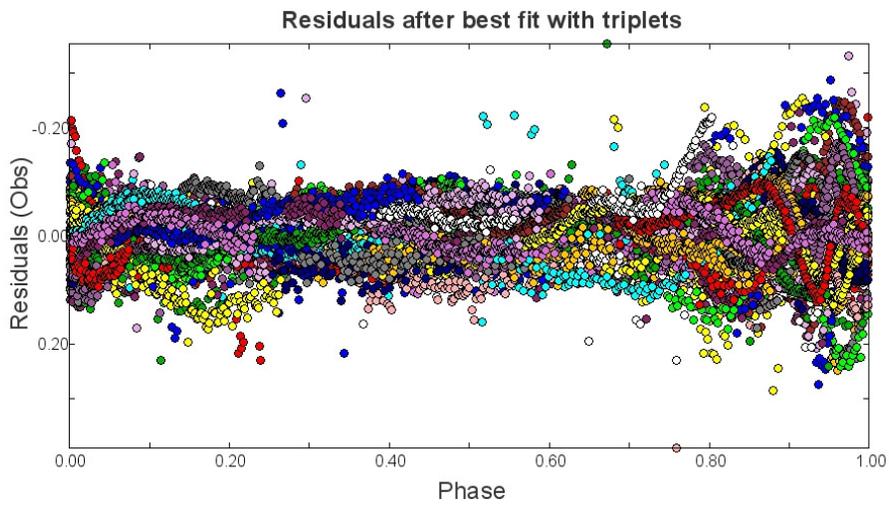

Figure 8a. 8b. Residuals after harmonics whitening and after triplet whitening

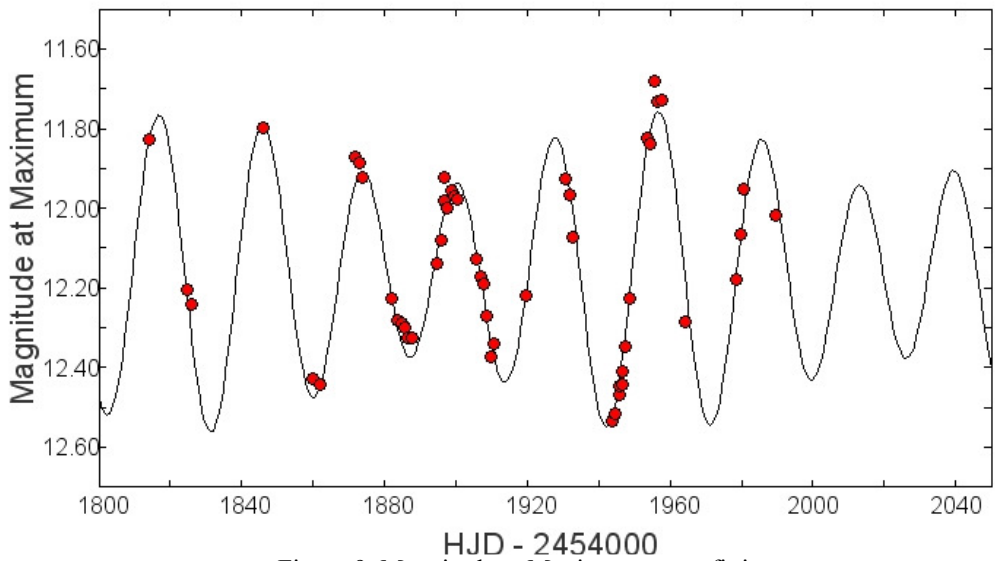

Figure 9. Magnitude at Maximum curve fitting

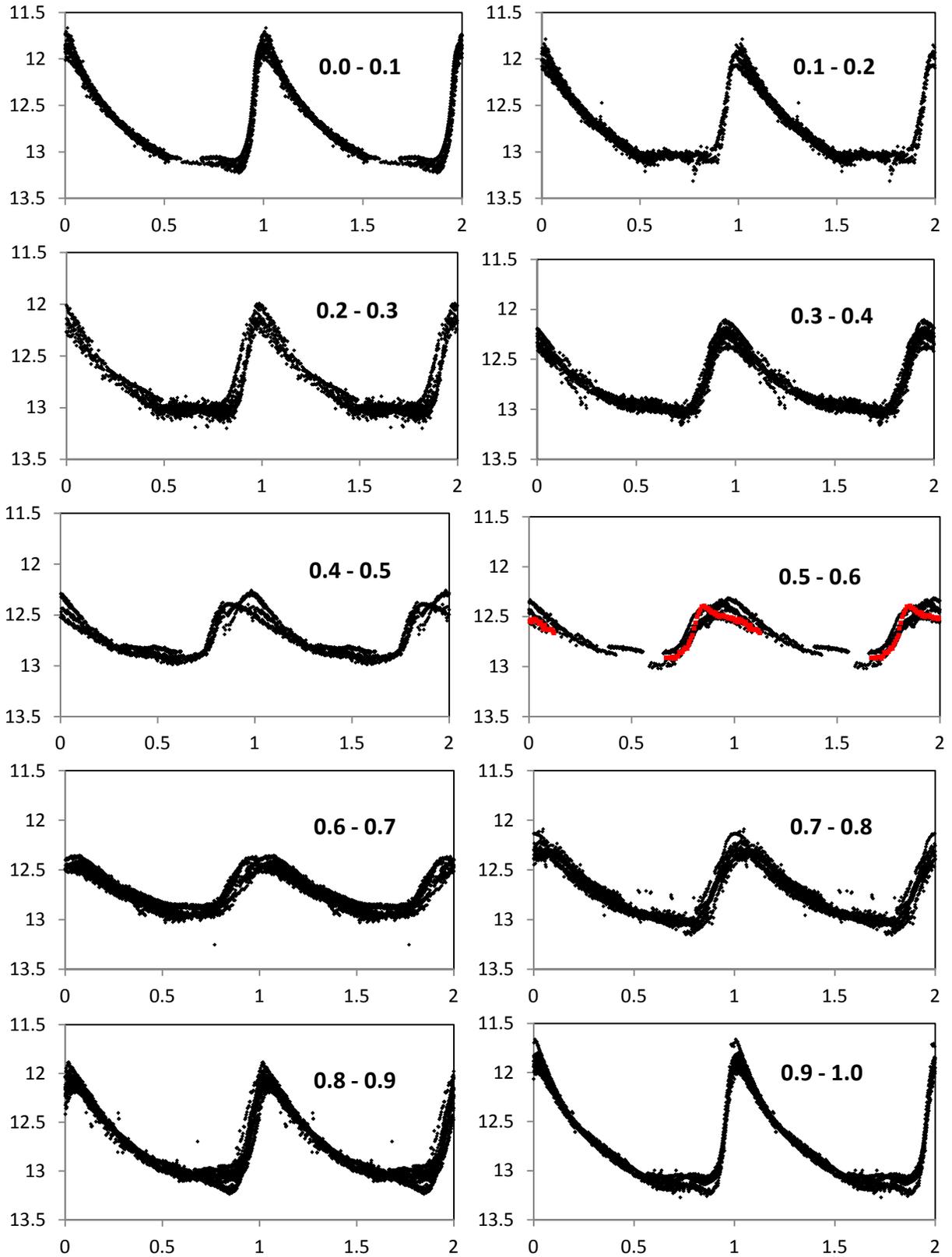

Figure 10. Light curve for different Blazhko subsets

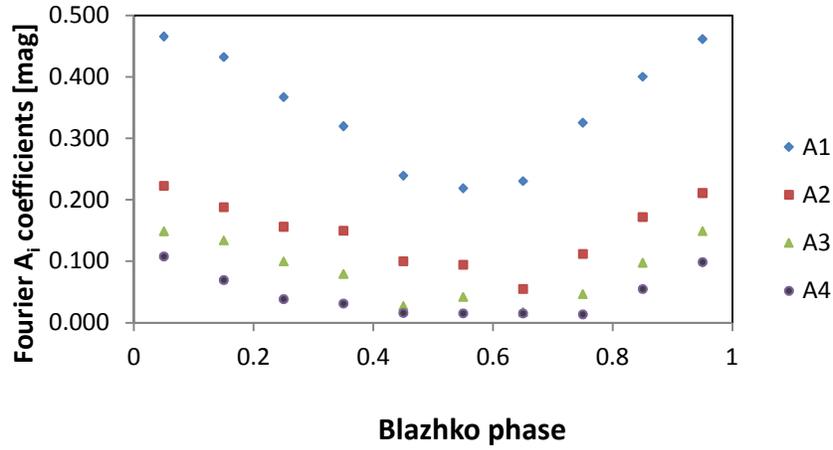

Figure 11a. Fourier $A_i$ amplitude [mag] variations versus Blazhko $\Psi_i$ subsets

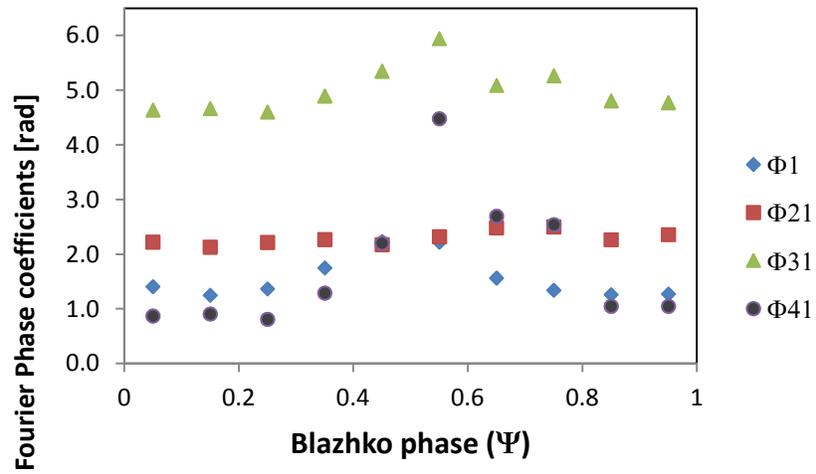

Figure 11b. Fourier $\Phi_1$ and $\Phi_{k1}$ phase [rad] variations versus Blazhko $\Psi_i$ subsets

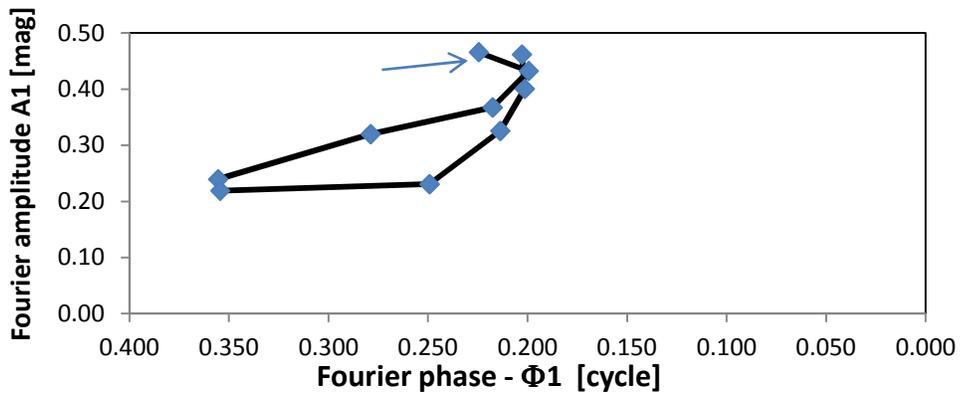

Figure 12, Fourier Amplitude $A_1$ versus Phase $\Phi_1$ for different Blazhko subsets.
*The point corresponding to the bin nearest to 0.0 Blazhko phase is indicated with a small arrow.*